\documentclass[11pt]{article}
\usepackage{graphicx}
\usepackage{amsmath}
\usepackage{amsfonts}
\usepackage{amssymb}
\newtheorem{theorem}{Theorem}

\newtheorem{remark}[theorem]{Remark}

\begin{document}

\author{\textbf{Scott M. Hitchcock}\\National Superconducting Cyclotron Laboratory (NSCL)\\Michigan State University, East Lansing, MI 48824-1321\\E-mail: hitchcock@nscl.msu.edu}
\title{\textbf{Is There a 'Conservation of Information Law' for the Universe?}}
\date{August 1, 2001}
\maketitle
\begin{abstract}
What are the implications if the total 'information' in the universe is
conserved? Black holes might be 'logic gates' recomputing the 'lost
information' from incoming 'signals' from outside their event horizons into
outgoing 'signals' representing evaporative or radiative decay 'products' of
the reconfiguration process of the black hole quantum logic 'gate'. Apparent
local imbalances in the information flow can be corrected by including the
effects of the coupling of the vacuum 'reservoir' of information as part of
the total information involved in any evolutionary process. In this way
perhaps the 'vacuum' computes the future of the observable universe.
\end{abstract}

\section{Introduction}

Let us begin with the assumption that there may exist a Conservation of Total
Information 'law' for the entire universe. This means that the total
information content in the current epoch is the same as that in the early
universe \cite{mmc2} regardless of the limitations on what we observe as the
'visible' forms it takes. The motivation for this is based in the idea of
conservation of total mass-energy for the universe regardless of the forms
matter takes during the reconfiguration processes of matter within the
framework of an expanding vacuum filled with growing quantum networks
\cite{qgn}.

If all current visible structures floating on the sea of the vacuum constitute
a very small percent of the total 'information' in the entire universe and the
'expansion' of 'space' combined with local gravitationally driven aggregation
of mass into 'information' sources and sinks (such as stars and planets for
instance) provide a means for 'computing' new configurations of matter
(biological systems for instance), then perhaps the remainder of the
'invisible information' is in the vacuum 'reservoir'. All unstable 'visible'
physical systems such as atoms and molecules represent the building blocks for
complex hierarchical systems.

If one were to take this view then the apparent information 'loss' by
'trapping' in black holes could be recast as the 'computation' of new forms of
information (Hawking radiation) by the black hole 'logic gate' in a quantum
computer space (vacuum). The signals emanating from the black hole carry
information content about the logical operations performed on the incoming
mass 'signals' contributing to the process of black hole formation. The black
hole recomputed its unstable state (coupled to the vacuum) into a more stable
one in which outgoing signals are 'emitted'. The analogy is similar to the
processing of incoming photons by the electrons around an atom into emission spectra.

Since new structures clearly emerge from previous 'unstable' configurations of
matter in the universe, new information also emerges as a function of the
entanglement of new combinations of quantum systems into a single system with
collective behaviors that are more than a linear sum (actually a 'direct
product') of the component systems by themselves. New information can be
created locally then but overall information may be conserved for the universe
as a whole due to the conversion of information from the expanding vacuum
reservoir into novel configuration information defining the islands of matter
throughout the universe.

\section{Black Hole Quantum Logic 'Gates'?}

Let us begin by looking at black holes as quantum computer logic gates in
which incoming matter (information) signals are recomputed into outgoing decay
product 'signals'. The 'difference' in the form and content of the
'information' between incoming and outgoing signals is the result of the
coupling of the black hole to the expanding space (vacuum) information
reservoir. In this sense the composite system of space and matter forms at
least a network of local quantum computers in the neighborhood of black holes.

The universe may then be a form of 'quantum computer' \cite{lloyd}
\emph{network} of black hole logic gates or other complex matter 'computers'
whose power to recompute the evolutionary progression of global states of the
expanding universe is due their use of the vacuum reservoir of information
created by the gravitational interaction of matter bound to the sea of vacuum
energy upon which they are small components.

Milan M. Cirkovic's idea (\cite{mmc}) that there may be no primordial black
holes (certainly believable in light of the absence of appropriate gamma ray
spectra) combined with the possibility that any black holes formed during the
evolution of the universe might not trap or evaporate enough 'information' at
a rate significant enough to create an imbalance in the total information of
the universe might be seen as a 'hint' that total information content in the
universe may to first order be conserved. Black holes are generally regarded
as information 'sinks' but they are still within the universe. While the
information is usually thought of as being 'lost' inside a BH, if we take the
view that the BH is a sort of 'memory' or logic gate connected with the rest
of the universe through its coupling to the vacuum (expanding space) then the
expanding universe is engaged in a 'race' between the 'creation',
'processing', 'transformation' and 'destruction' of 'information' encoded in
complex physical system 'islands' in the vacuum info-space. The rate of
expansion (possibly accelerating according to recent supernovae observations)
can be used to test whether a conservation of information law is valid.

The information processing time (lifetime?) of a black hole seems to be
dependent on the information (e.g. consumed 'signals' plus remnant mass of its
stellar progenitor) that it holds. Even if this information is an entangled
'collective excitation' mess, it can decohere (by a vacuum-BH interaction that
triggers the 'output' of decay products?) via the vacuum induced evaporation
of the BH at a critical transition 'mass'. The key to accounting for all the
information is to remember that the black hole exists only because there is an
environment (the vacuum) that defines it.

Black holes represent one of the fundamental testing grounds for a
conservation of information law for the entire universe. Any unstable system
presents the opportunity to test conservation of information. Black holes and
the interest in irreversible loss of information seems to have caught the
imagination of cosmologists and those interested in quantum processes in which
information is encoded in the form of 'observable' properties of 'physical'
systems. There is a possibility that the 'entropy' issues can be clarified
when considering the 'remainder' of information shifted into the vacuum
'information reservoir' after processing of the 'lost' signals (e.g. signal
trapping and mass accretion resulting in creation of at least one 'signal' in
the form of the 'inflation of event horizon', etc.) into outgoing signals
during the evaporation or decay process. Gravity is a sort of adaptive
'wiring' between BHs and matter outside the event horizon.

Black holes viewed as computational logic gates that recompute gravitationally
wired signals into new forms of information provide a logical source for
Hawking radiation and other possible 'evaporation' signals. This perspective
might illuminate the features of a general information conservation law that
can be used to establish how complex systems can arise as the result of the
computational effects of expanding vacuum upon the matter within it.

\subsection{Info-flow through BH Gates}

The state of the entire universe, $\left|  U\right\rangle $, is the direct
product its entangled component sub-states including the vacuum $\left|
V_{U}\right\rangle $, the physical (observable) 'matter' sub-systems $\left|
S_{U}\right\rangle $ (such as particles, nuclei, atoms, molecules and
gravitational aggregations of these in planets, stars, galaxies and
life-forms), 'signals' $\left|  \lambda_{U}\right\rangle $(e.g. photons)and
the expansion 'boundary condition' or 'surface area' of the 'event horizon' of
the expanding vacuum energy density characterized by the expansion front
bubble surface area, $\left|  A_{U}\right\rangle $:%

\begin{equation}
\left|  U\right\rangle {\Large =}\left|  V_{U}\right\rangle \bigotimes\left|
S_{U}\right\rangle \bigotimes\left|  \lambda_{U}\right\rangle \bigotimes
\left|  A_{U}\right\rangle
\end{equation}

Which becomes:%

\begin{equation}
\left|  U\right\rangle {\Large =}\left|  V_{U}\right\rangle \bigotimes\left[
{\displaystyle\bigotimes\limits_{i=1}^{\infty}}
{\LARGE \left|  S_{i}\right\rangle }\right]  \bigotimes\left[
{\displaystyle\bigotimes\limits_{j=1}^{\infty}}
{\LARGE \left|  \lambda_{j}\right\rangle }\right]  \bigotimes\left|
A_{U}\right\rangle
\end{equation}

Where causal networks can be formed by destabilized physical sub-systems and
the signals between them in a general environment of an expanding vacuum
'information' reservoir bounded by a computational enclosure of the expansion
front metricized by the Hubble flow and large scale dynamics of galaxies. We
note here that the 'vacuum' is not empty, but to the contrary, contains most
of the information in the universe. The observable matter we consider to be
the sources of information about the large scale structure of the universe are
relic computational perturbations from the anisotropic phonon-like collective
excitation decay modes of the inflationary epoch. The Planck scale relics were
'computed' into the particles we are built by the conversion of the localized
energy of the Planck epoch universe into 'decay products' like matter and the vacuum.

If we look at information flow through a single black hole logic gate,
$\left|  BH\right\rangle $, we see that an incoming signal, $\left|
\lambda_{IN}\right\rangle $, 'lost' by entering a BH gate is computed into to
new signals such as the quantized increase in the area of the event horizon,
and a quantized increase in the total mass of the BH. If the BH mass is below
the critical excited state at which quantum evaporation processes result in
signal emissions, then the BH act like a long term RAM storage device or
memory for the incoming signals delivered by the gravitational 'wiring' of the
vacuum to external sub-systems of the universe. At this point total
information is still conserved and the 'lost' information embedded in the BH
has been computed into observable changes in the collective excitation states
such as total BH mass and event area.

The state of the BH subsystem, $\left|  S_{BH}\right\rangle $, encompassing
the space in which incoming signals make a transition from 'classical'
separation (i.e. superposition) to quantum coupling (detection, absorption, or
entanglement) before gravitational detection of a signal, ignoring for the
moment the expansion area boundary conditions, $\left|  A_{U}\right\rangle $,
for the rest of the universe, is a 'classical' quantum system:%

\begin{equation}
\left|  S_{BH}\right\rangle _{C}{\Large =}\left|  BH_{0}\right\rangle
{\Large +}\left|  \lambda_{IN}\right\rangle
\end{equation}

Upon detection of the signal by gravitational trapping of the 'quantum' state
of the BH system, $\left|  S_{BH}\right\rangle _{Q}$, and the 'excited' state,
$\left|  BH^{\ast}\right\rangle $, of the BH information processing system is
a composite system of the entangled signal, $\left|  \lambda_{IN}\right\rangle
$, with the BH 'mass' inside its now expanded event horizon. The increase in
the event horizon area is $\left|  \delta A_{BH}\right\rangle $, and therefore
information content of the BH is characterized by $\left|  BH^{\ast
}\right\rangle $:%

\begin{equation}
\left|  S_{BH}\right\rangle _{Q}{\Large =}\left|  M_{BH}\right\rangle
\bigotimes\left|  \lambda_{IN}\right\rangle =\left|  M_{BH}\right\rangle
\bigotimes\left|  \delta A_{BH}\right\rangle =\left|  BH^{\ast}\right\rangle
\end{equation}

The decay of this state occurs only if it is the critical evaporation
threshold state, $\left|  S_{BH}\right\rangle _{Critical}$. The system can
gravitationally detect a 'less than infinite' number of signals in a series of
hierarchical excited states corresponding to the increased mass and
computational (event horizon) surface area before it reaches this state in
which the gate converts the event horizon and accretion mass information as
Hawking evaporation signals. If the BH system is at a point where the emission
of signals by 'evaporation' occurs then the coupling of the vacuum to the
remnant links the differences in the form and content of the pre-BH processed
signals to the emitted post computation signals.

Since all forms of information during the process of 'loss' and delayed
'evaporation' in black holes (including any entropy terms that are in fact
communicated to the vacuum information reservoir) can be accounted for, we see
that black holes obey conservation of total information when all system
components and environment involved in the reconfiguration of information from
input to output are taken into account.

This limiting case of conservation of information for a black hole illustrates
that a conservation of information law for the entire universe may provide a
convenient tool for understanding fundamental processes, the evolution of
complex systems and cosmological effects in local 'matter' islands.

\begin{remark}
Note that this process pertains also to non-BH physical systems that detect
and process information such as Feynman Cocks (FCs), Collective Excitation
Networks (CENs) and Sequential Excitation Networks (SENs) \cite{hitchcock},
\cite{hitchcock2}, \cite{hitchcock3}.
\end{remark}

I would like to thank Paola A. Zizzi and Milan M. Cirkovic for inspiration to
pursue this speculative direction. They are both blameless for my errors.
Special thanks to poolside Toni. For further information and papers see: http://www.nscl.msu.edu/departments/facilities/hitchcock/index.htm


\begin{thebibliography}{9}
\bibitem{mmc2}''\textbf{Is the Universe Really so Simple?'' }by \emph{Milan M.
Cirkovic}, LANL e-print archives quant-ph/0107070, 13 Jul 2001. http://arxiv.org/ftp/quant-ph/papers/0107/0107070.pdf

\bibitem{qgn}\textbf{''The Early Universe as a Quantum Growing Network'' }by
Paola A. Zizzi, LANL e-print archives, gr-qc/0103002 v3, 6 Apr 2001, http://arxiv.org/ftp/gr-qc/papers/0103/0103002.pdf

\bibitem{lloyd}\textbf{''Universe as quantum computer''} by \emph{Seth Lloyd},
LANL e-print archives, quant-ph/9912088, 17 Dec 1999.

\bibitem{mmc}Private communication with Milan M. Cirkovic, Astronomical
Observatory Belgrade, Yugoslavia dated August 1, 2001.

\bibitem{hitchcock}''\textbf{Quantum Clocks and the Origin of Time in Complex
Systems}'' by \emph{Scott Hitchcock}, LANL e-print archives; gr-qc/9902046 v2,
20 Feb 1999, also NSCL Publication: MSUCL-1123, 1999. Available at http://www.nscl.msu.edu/news/nscl\_library/nscl\_preprint/MSUCL1123.pdf

\bibitem{hitchcock2}''\textbf{Feynman Clocks, Causal Networks, and the Origin
of Hierarchical 'Arrows of Time' in Complex Systems from the Big Bang to the
Brain. Part I 'Conjectures'''}by \emph{Scott Hitchcock}, LANL e-print
archives; gr-qc/0005074, 16 May 2000, also NSCL Publication: MSUCL-1135, 2000.
Available at http://www.nscl.msu.edu/news/nscl\_library/nscl\_preprint/MSUCL1135.pdf

\bibitem{hitchcock3}\textbf{''Feynman Clocks, Causal Networks, and
Hierarchical Arrows of Time in Complex Systems From the Big Bang to the
Brain''} An invited talk and paper given by \emph{Scott Hitchcock }at the
'XXIII International Workshop on the Fundamental Problems of High Energy
Physics and Field Theory' at the Institute for High Energy Physics (IHEP),
Protvino, Russia, June 21-23, 2000. To be published in the Proceedings of the
Workshop and as NSCL as pre-print (MSUCL -1172). It is also at the Los Alamos
National Labs, LANL e-Print Archives report number quant-ph/00100014.
Available at http://www.nscl.msu.edu/news/nscl\_library/nscl\_preprint/MSUCL1172.pdf
\end{thebibliography}
\end{document}